\documentclass[anonymous=false]{acmart}

\usepackage[utf8]{inputenc}
\usepackage{multirow}
\usepackage{xcolor,colortbl} 
\usepackage{wrapfig}
\usepackage{subcaption}
\usepackage{natbib}
\usepackage{graphicx} 

\usepackage[textwidth=1.5cm]{todonotes}
\usepackage{markdown}

\copyrightyear{2024}
\setcopyright{none}

\begin{document}

\title{Data Generation via Latent Factor Simulation for Fairness-aware Re-ranking}

\author[E. Stefancova]{Elena Stefancova}
\email{elena.stefancova@fmph.uniba.sk}
\affiliation{%
  \institution{Comenius University Bratislava}
  \streetaddress{}
  \city{Bratislava}
  \country{Slovakia}
}

\author[C. All]{Cassidy All}
\email{cassidy.all@colorado.edu}
\affiliation{%
  \institution{Department of Information Science; University of Colorado, Boulder}
  \streetaddress{}
  \city{Boulder}
  \state{Colorado}
  \country{USA}
  \postcode{80309}
}

\author[J. Paup]{Joshua Paup}
\email{joshua.paup@colorado.edu}
\affiliation{%
  \institution{Department of Information Science; University of Colorado, Boulder}
  \streetaddress{}
  \city{Boulder}
  \state{Colorado}
  \country{USA}
  \postcode{80309}
}

\author[M. Homola]{Martin Homola}
\email{homola@fmph.uniba.sk}
\affiliation{%
  \institution{Comenius University Bratislava}
  \streetaddress{}
  \city{Bratislava}
  \country{Slovakia}
}

\author[N. Mattei]{Nicholas Mattei}
\email{nsmattei@tulane.edu}
\affiliation{%
  \institution{Department of Computer Science; Tulane University}
  \streetaddress{}
  \city{New Orleans}
  \state{Louisiana}
  \country{USA}
  \postcode{70118}
}

\author[R. Burke]{Robin Burke}
\email{robin.burke@colorado.edu}
\orcid{0000-0001-5766-6434}
\affiliation{%
  \institution{Department of Information Science; University of Colorado, Boulder}
  \streetaddress{}
  \city{Boulder}
  \state{Colorado}
  \country{USA}
  \postcode{80309}
}

\begin{abstract}
Synthetic data is a useful resource for algorithmic research. It allows for the evaluation of systems under a range of conditions that might be difficult to achieve in real world settings. In recommender systems, the use of synthetic data is somewhat limited; some work has concentrated on building user-item interaction data at large scale. We believe that fairness-aware recommendation research can benefit from simulated data as it allows the study of protected groups and their interactions without depending on sensitive data that needs privacy protection. In this paper, we propose a novel type of data for fairness-aware recommendation: synthetic recommender system outputs that can be used to study re-ranking algorithms. 
\end{abstract}



\maketitle

\section{Introduction}
Research in fairness-aware recommendation usually takes two paths: model-based, in which fairness objectives are integrated into the machine learning model itself, and post-processing or re-ranking, in which fairness objectives are applied to re-order the results of a recommender system that is not itself fairness-aware. The third option, pre-processing the input data to support fairness is less often applied in recommender systems. In this paper, we concentrate on supporting fairness-aware recommendation research that uses a post-processing approach.

Studying fairness-aware post-processing requires as input pre-computed recommendations that can be re-ranked and their fairness properties studied. While it is relatively straightforward to use existing recommendation data sets and frameworks to produce recommendations for this purpose, the set of available data sets with protected features around which fairness concerns might arise is relatively limited. Researchers often use a relatively limited repertoire of such data sets. In other cases, researchers may invent fairness concerns independent of detailed analysis of the fairness properties of the domain; for example, by designating some percentage of least popular items as protected. We believe such methodologies are valid and valuable (and part of our own experimental practice). However, in all such cases, researchers are limited by the characteristics of the data set and induced recommendations including its particular distribution of protected features. 

In this paper, we describe our methodology for creating synthetic recommendation lists for studying fairness-aware re-ranking: LAtent Factor Simulation. As the name implies, we create synthetic data via first simulating the latent factor matrices that a matrix factorization model could produce and then generate sample ratings from these matrices. We show that it is possible to produce recommendation lists
with characteristics similar to those that come from factorization models, and that we can adjust a number of dataset characteristics related to protected groups to evaluate fairness-aware re-rankers under a variety of conditions. 

\section{Related Work}
There have been a number of existing approaches to creating synthetic data for recommender systems. The first category is generation methods that use heuristics to generate data. These approaches leverage strong assumptions on the dataset distributions and properties.  DataGenCARS \citep{del2017datagencars} uses a description of user profiles to generate a dataset with specified properties. The designer of the algorithm manually designs and describes user ``profiles''. Another approach is to mine generic profiles from historical data by user clustering, as done in \citep{monti2019all}, and these profiles can be extended to include additional features and contexts as in \citep{pasinato2013generating}. In these works, the experimenter has to manually craft the user model, and make assumptions on how users behave. The quality of the resulting data will only be as good as these assumptions, however, and it is difficult to capture the texture of user-item interactions in their full detail.

An alternative to a manually-specified model is one that is learned from a data source. In unpublished work, Lin used a traditional matrix factorization model, and then used variational auto-encoders to generate new user vectors~\cite{lin2020synthetic}. Researchers in reinforcement learning often use synthetic data generation for experimentation \citep{zou2019reinforcement,zhou2020interactive}, including latent vector models. However, these models are often unrealistic, or too simple. For instance, authors in \citep{friedman2016differential} use 20-dimensional user latent vectors, which is very low (typical values in recommender systems range between 100 and 200). Authors of \citep{zhou2020interactive} generate a dataset that has 40\% sparsity (40\% of all possible combinations of users and items contain ratings), whereas industrial-grade datasets have sparsity orders of magnitude less (\~0.1\%). A method for generating realistically sparse datasets for recommendation is described in \cite{belletti2019scalable}. However, the Kronecker product expansion described here, a graph theoretic technique, creates both synthetic users and synthetic items, and does not generate item or user metadata, so cannot for studying fairness-aware recommendation.

Another approach to creating data that can be shared without compromising user privacy is that of data obfuscation. For example, \citep{weinsberg2012blurme} and \citep{slokom2019data} are examples of algorithms that use this approach. However, because the ratings matrix is modified but not replaced, each entry corresponds to a real user who identity might be compromised. Because LAFS is completely synthetic, no similar risks exist. A probabilistic approach to handling protected features was explored in \cite{burke2018synthetic}, which used rating data properties to derive behavioral features to stand in for (unknown) demographic data.  

As noted above, we concentrate in this paper on supporting research in fairness-oriented re-ranking. This approach represents a substantial and on-going research arc in fairness-aware recommendation. One of the advantages of post-processing is that the fairness objective can be adapted on the fly without relearning the recommendation model. It is also the case that specific guarantees relative to item distribution can be ensured by conducting batch reranking across an entire set of recommendations to be delivered. Such guarantees are not available if when fairness is included in part of the training objective. Re-ranking is also advantageous when the tradeoff between fairness objectives in inherently dynamic as in \cite{aird2024dynamic}. Re-ranking has been used to address a wide variety of fairness-aware recommendation challenges. There has been considerably investigation of greedy sublist optimization as a technique. In binary fairness situations, we note the algorithms in \citep{Celis2018-zh,Ekstrand2021-iu,10.1109/TKDE.2011.15,Gomez2021-av,Garcia-Soriano2021-gx,liu2018personalizing,sonboli2020opportunistic,aird2024dynamic}. The other main category of re-ranking solutions are batch-oriented approaches that consider the whole collection of recommendation lists at once and optimize across all lists aimed at all users. Examples of this technique include \citep{biega2018equity,Singh2018-zy,surer2018multistakeholder}. A detailed survey of approaches to fairness-aware recommendation can be found in \cite{ekstrand2022fairness}.

\section{Latent Factor Simulation}

\begin{figure}
    \centering
    \includegraphics[width=0.60\linewidth]{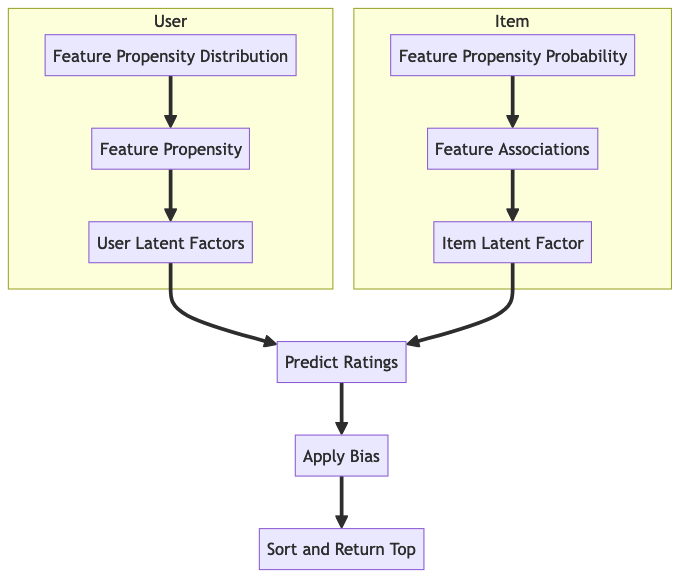}
    \Description[A graphic showing the multiple steps in data generation]{These are the same steps described below.}
    \caption{Overview of the LAFS data generation process}
    \label{fig:lafs-process}
\end{figure}

The LAFS data generation process happens in several stages. See Figure~\ref{fig:lafs-process}.

\subsection{Latent Factors}


\begin{center}
\begin{table}[tbh]
{\small
    \begin{tabular}{|c|c|}
    \hline
    $n_i$ & The number of items \\
    $n_u$ & The number of users per propensity distribution \\
    $k$ & The number of latent factors \\
    $s$ & The number of sensitive factors ($s < k$) \\
    $\sigma_f$ & Standard deviation for factor generation \\
    \hline
    $d_u = [(\mu_{u1}, \sigma_{u1}), (\mu_{u2}, \sigma_{u2})..., (\mu_{uk}, \sigma_{uk})] $ & Feature propensity distributions for users \\
    $\Pi_u = [\pi_{u1}, \pi_{u2}, ..., \pi_{uf}]$ & Feature propensities for users \\
    $\Pi_U = [\Pi_u \forall u]$ & The matrix of all user-feature associations \\
    $U$ & $<n_u \times n_f>$ matrix of user latent factors \\
    \hline
    $d_i = [d_{i1}, d_{i2},... , (d_{if}] $ & Feature propensity probabilities for items \\
    $\Pi_i = [\pi_{i1}, \pi_{i2}, ..., \pi_{if}]$ & Feature associations for item $i$ \\
    $\Pi_I = [\Pi_i \forall i]$ & The matrix of all item-feature associations \\
    $V$ & $<n_i \times n_f>$ matrix of item latent factors \\
    \hline
    $l'$ & The size of the initial recommendation list \\
    $l$ & The size of the output recommendation list \\
    $B = [(\mu_1, \sigma_1), (\mu_2, \sigma_2), ... (\mu_f, \sigma_s)] $ & Bias generators for sensitive features \\
    \hline
    \end{tabular}
}
\caption{Notation for LAFS data generation}
\label{tab:notation}
\end{table}
\end{center}

As shown in Table~\ref{tab:notation}, $U$ and  $V$ are the user and item latent factor matrices with $k$ latent factors. We designate the first $k_s$ of the latent factors as corresponding to protected features of items, and the remaining $k - k_s$ factors correspond to other aspects of the items. Thus, we can generate data with multiple protected features for studying intersection impacts of fairness-aware re-ranking.

We generate these latent factors for each (simulated) user and item in a two-step process. First, we generate two matrices of propensities that has a value for each user and each factor, and we do the same for each item and each factor. Then, from the propensities we generate a latent factor value for the corresponding $U$ or $V$ matrix. This two-step process avoids having the latent factors tied exactly to the user propensities, which would reduce the variance of the generated factors and make the recommendation lists too similar to each other.

More formally, the experimenter specifies a vector of $d_i$ values, one for each factor, representing the probability of an item be associated with that factor. The assumption is that latent factors are closely tied to the presence or absence of a feature for each item and so each $\pi_i$ value is binary: the item has the feature or it doesn't. For each item $i$ and factor $j$, each binary $pi_{ij}$ propensity value is computed by treating the $d_i$ value as a probability in a Bernoulli trial. 

This binary propensity label for items has the additional advantage that it is easy to identify items as protected relative to the $k_s$ sensitive features. We can control how many items will be protected relative to different sensitive by changing the $d_i$ probabilities for these sensitive features.

Once we have the $\Pi_I$ propensity matrix, we generate the item latent feature matrix $V$ by sampling from a normal distribution centered around the corresponding propensity. That is, $v_{ij} = \mathcal{N}(\pi_{ij}, \sigma_f)$.

For user latent factors, the process is slightly more complex because we do not assume binary associations between users and latent factors. Instead of specifying the probability of the association for a Bernoulli trial, the experimenter specifies feature-specific parameters for a normal distribution: $(\mu_{u1}, \sigma_{u1})$. The $Pi_U$ matrix is filled by drawing from this distribution, and then, as with the $V$ matrix, the $U$ is filled by drawing from a normal distribution centered on the propensity: $u_{uj} = \mathcal{N}(\pi_{uj}, \sigma_f)$. This can be performed for users groups of different sizes and propensity distributions.

The latent factors for users and items are therefore created through similar two-step processes. First, propensities are drawn from experimenter-specified distributions. Second, latent factor values are drawn from a normal distribution centered on the prospensities.

Note that we do not require experimenters to specify the full co-variance matrix for their distributions. We assume that the propsensities and latent factor are independent. The values of the latent factors are not constrained and do not have to be in any particular range, similar to latent factors derived from unconstrained matrix factorization.

\subsection{Rating Generation}

Once we have the latent factor matrices $U$ and $V$, the next step is to generate recommendation lists from them. For each user $u$, we select a set of $l'$ items at random. For each item, we compute the rating by multiplying the $U_u$ and $I_i$ latent factor vectors. 

To simulate bias against protected group items, we apply a randomly generated bias penalty (which could be zero) to the computed rating for each sensitive feature that it possesses. The penalty is drawn from an experiment-specified vector of bias distributions: $B$. 

The final set of items and their adjusted ratings are then sorted and the top $l$ items returned as the simulated recommendation list. This sorting process simulates the sorting that occurs when a recommender system is producing top-k recommendations for the user. We would expect that recommendation lists would contain items with higher rating predictions.

After all recommendations have been generated, an experimenter-specified min-max transformation is applied to normalize the scores from all of the generated recommendations ($l'$ lists) to the experimenter's preferred range. This is not strictly necessary but helps make the recommendation output easier to understand. 

\subsection{User dynamics}

For some of the experiments that we are conducting, we are interested in the dynamic properties of a recommender system with regard to fairness and in particular, how fairness outcomes may shift when the user population shifts. To support this kind of experiment, we have added an additional feature to LAFS, which allows for a sequence of user \textit{regimes} in the recommendation output. A user regime is defined by a particular $d_u$, that is user feature propensity distribution. We wish to start with a user base that is very interested in items containing the protected feature and then move to a user base that is not. 

To support these types of chronological shifts, we allow for a list of $d_u$ distributions $[d_u^1, d_u^2, ..., d_u^r]$ where $r$ is the desired number of regimes and a list of $n_u$ user counts $[n_u^1, n_u^2, ... n_u^r]$. To generate the recommendation data, we iterate through the list of distributions, producing $n_u^1$ users using the distribution $d_u^1$ and then $n_u^2$ users with the next distribution, and so on, until all the users are generated. 

The code for LAFS is available as open source on GitHub under the MIT License\footnote{https://github.com/that-recsys-lab/lafs}.

\section{Conclusion}

One thing that has been clear in this work is that there is a lack of an established methodology for evaluating data set simulation in recommender systems. Part of the reason is that, beyond basic measures like sparsity, we do not have strong predictive measures connecting data set properties with recommender system performance and it is therefore difficult to assess whether findings relative to a synthetic data set will be predictive of performance on a real-world one. 

Clearly, additional research is needed on this and other aspects of simulated evaluation of recommender systems. As we continue to work with LAFS, we plan to explore a variety of visualizations, metrics and other techniques to compare its simulated output with recommendations from real algorithms and also to compare the synthetic latent factors with those derived from real-world data sets. 

While the LAFS generation method has proved useful for our purposes (see \cite{aird2024dynamic}), there are additional improvements that we envision. Currently, we do not represent item popularity and so items in user's initial recommendation lists are selected uniformly at random.It would be more realistic to assign items to ranks along a long-tail distribution to better mimic the distribution of items in recommendation lists. Also, we have found that, in some data sets, there is correlation (and anti-correlation) between sensitive features, so, rather than assuming independence, it may be useful to allow experimenters to specify the co-variance matrix for these features. We plan to include these features in our next release of the tool.



\section*{Acknowledgements}
This research was supported by the National Science Foundation under awards IIS-2107577 and IIS-2107505. This work was also supported by the Tatra Bank Foundation, and the national projects nos. VEGA-1/0621/22 and APVV-20-0353 awarded by Slovak VEGA and APVV research agencies.

\bibliographystyle{ACM-Reference-Format}
\bibliography{scruf}

\end{document}